\documentclass[aps,prb,a4paper,superscriptaddress,twocolumn,floatfix,showpacs]{revtex4}
\usepackage{graphicx}
\usepackage{latexsym}
\usepackage[]{amsmath}

\begin{document}
\newcommand{\chem}[1]{\ensuremath{\mathrm{#1}}}

\title{Two-dimensional magnetism in the pnictide superconductor parent material \\ 
SrFeAsF probed by muon-spin relaxation}

\author{P.\ J.\ Baker}
\affiliation{Oxford University Department of Physics, Clarendon Laboratory,
Parks Road, Oxford OX1 3PU, United Kingdom}

\author{I.\ Franke}
\affiliation{Oxford University Department of Physics, Clarendon Laboratory,
Parks Road, Oxford OX1 3PU, United Kingdom}

\author{T.\ Lancaster}
\affiliation{Oxford University Department of Physics, Clarendon Laboratory,
Parks Road, Oxford OX1 3PU, United Kingdom}

\author{S.\ J.\ Blundell}
\affiliation{Oxford University Department of Physics, Clarendon Laboratory,
Parks Road, Oxford OX1 3PU, United Kingdom}


\author{L.\ Kerslake}
\affiliation{Department of Chemistry, University of Oxford, Inorganic 
Chemistry Laboratory, South Parks Road, Oxford, OX1 3QR, United Kingdom}



\author{S.\ J.\ Clarke}
\affiliation{Department of Chemistry, University of Oxford, Inorganic 
Chemistry Laboratory, South Parks Road, Oxford, OX1 3QR, United Kingdom}

\date{\today}

\begin{abstract}
We report muon-spin relaxation measurements on SrFeAsF, which is the parent 
compound of a newly discovered iron-arsenic-fluoride based series of 
superconducting materials. 
We find that this material has very similar magnetic properties to LaFeAsO, 
such as separated magnetic and structural transitions ($T_{\rm N} = 120$~K, 
$T_s = 175$~K), contrasting with SrFe$_2$As$_2$ where they are coincident.
The muon oscillation frequencies fall away very sharply at $T_{\rm N}$, which 
suggests that the magnetic exchange between the layers is weaker than in 
comparable oxypnictide compounds. This is consistent with our specific heat 
measurements, which find that the entropy change $\Delta S = 0.05$~Jmol$^{-1}$K$^{-1}$ 
largely occurs at the structural transition and there is no anomaly at $T_{\rm N}$.
\end{abstract}


\maketitle

Quasi-two-dimensional magnets on square lattices are the subject of 
considerable theoretical and experimental 
attention.~\cite{manousakis91,kastner98,sengupta03} This has primarily 
been due to the success of models of the spin-$1/2$ Heisenberg antiferromagnet 
in describing the physics of \chem{La_{2}CuO_{4}}, which is the 
prototypical parent compound of high-$T_c$ cuprate superconductors. 
\chem{La_{2}CuO_{4}} shows a tetragonal to orthorhombic structural transition 
at $T_o \simeq 530$~K and N\'{e}el ordering at $T_{\rm N} \simeq 325$~K. 
That $T_{\rm N}$ is far smaller than the antiferromagnetic exchange constant 
$J \sim 1500$~K demonstrates that this compound has a remarkably large magnetic 
anisotropy, with weak coupling between the \chem{CuO_{2}} 
layers.~\cite{kastner98,yasuda05} The magnetic parent compounds of FeAs-based 
superconductors such as LaFeAsO$_{1-x}$F$_x$~\cite{kamihara08} have Fe atoms on a layered 
square lattice, and it is interesting to note that, like \chem{La_{2}CuO_{4}}, 
these have a tetragonal to orthorhombic structural distortion followed by 
antiferromagnetic ordering (e.g. Ref.~\onlinecite{zhao08nmat}). Here we study 
the magnetic properties of a newly discovered parent compound to a series of 
fluoropnictide superconductors, SrFeAsF,~\cite{tegel08arxiv,han08prb} where the 
fluoride ions should provide weaker magnetic exchange pathways between the FeAs 
layers than for {\it Ln}FeAsO or {\it A}Fe$_2$As$_2$ compounds.

Doped fluoropnictide compounds based on CaFeAsF and SrFeAsF have recently been found to 
superconduct,~\cite{matsuishi08jacs,wu08arxiv,matsuishi08jpsj,matsuishi08arxiv,zhu08arxiv} 
with comparable transition temperatures to the previously discovered oxypnictide 
compounds based on {\it Ln}FeAsO. These have similar FeAs layers to the 
oxypnictides, but divalent metal - fluoride layers replace the rare-earth - oxide 
layers.
Fluoropnictides can be doped on the Fe site, as for \chem{CaFe_{0.9}Co_{0.1}AsF} 
($T_c = 22$~K),~\cite{matsuishi08jacs} or the divalent metal site, 
as for Sr$_{0.5}$Sm$_{0.5}$FeAsF ($T_c = 56$~K),~\cite{wu08arxiv} and 
several approaches have already been 
explored.~\cite{matsuishi08jacs,wu08arxiv,matsuishi08jpsj,matsuishi08arxiv,zhu08arxiv}
The magnetic, electronic, and structural properties of the parent compounds 
CaFeAsF and SrFeAsF, and also EuFeAsF, have already been investigated. 
All three show transitions evident in resistivity and dc magnetization 
measurements, at $T_s = 120$, $175$, and $155$~K 
respectively.~\cite{matsuishi08jacs,matsuishi08jpsj,zhu08arxiv,tegel08arxiv,han08prb}
In SrFeAsF the structural transition has been probed using X-ray diffraction, 
and changes in the magnetism using M\"{o}ssbauer spectroscopy.~\cite{tegel08arxiv} 
The structural change varies smoothly below 175~K whereas the M\"{o}ssbauer 
spectra became increasingly complicated as the temperature is reduced. It is 
also interesting that the sign of the Hall coefficient $R_{\rm H}$ in SrFeAsF 
is reported to be positive below $T_s$, whereas it is negative in the undoped 
LaFeAsO and BaFe$_2$As$_2$ parent compounds.~\cite{han08prb} This could result 
from a different electronic structure near the Fermi surface, which might have 
implications for both the magnetism of the undoped compound and the 
superconductivity that emerges when it is doped.

The magnetism of {\it Ln}FeAsO compounds has already been intensively 
investigated by a wide range of techniques. 
Neutron diffraction measurements have been carried out on some of the 
undoped oxypnictides: LaFeAsO,~\cite{delacruz08nature} NdFeAsO,~\cite{chen08prb} 
PrFeAsO,~\cite{kimber08prb} and CeFeAsO.~\cite{zhao08nmat}
The results in each case indicate similar structural transitions 
at around $T_s = 150$~K, followed by long range, three dimensional 
antiferromagnetic ordering of the iron spins with significantly reduced 
moments $<1~\mu_{\rm B}$/Fe at $T_{\rm N}$, around 20~K below $T_s$, 
confirmed by other techniques.~\cite{mcguire08prb,mcguire08njp}
These features move to lower temperature with increasing doping 
and are absent in the superconducting phase for 
LaFeAsO$_{1-x}$F$_x$~\cite{luetkens08arxiv} and 
CeFeAsO$_{1-x}$F$_x$,~\cite{zhao08nmat} although magnetism and superconductivity 
seem to coexist over a small doping range in SmFeAsO$_{1-x}$F$_x$.~\cite{drew08arxiv}
In constrast, SrFe$_2$As$_2$ has coincident magnetic and structural ordering 
occurring in a first-order phase transition at $T_o = 205$~K.~\cite{krellner08prb} 
It seems that in general $A$Fe$_2$As$_2$ materials have more closely 
related structural and magnetic phase transitions, and more three-dimensional 
magnetism than the single layer FeAs materials.
With the discovery of new fluoroarsenide parent materials it is important to 
compare the magnetic structures and the separation between $T_s$ and $T_{\rm N}$ 
in the oxide-arsenide and fluoro-arsenide materials. 
Here we address these comparisons in SrFeAsF using the techniques of muon-spin 
relaxation, which is a local probe of the magnetic fields inside the sample and 
their dynamics, and also specific heat measurements which examine the changes in 
entropy at the transitions. 

\begin{figure}[t]
\includegraphics[width=8.5cm]{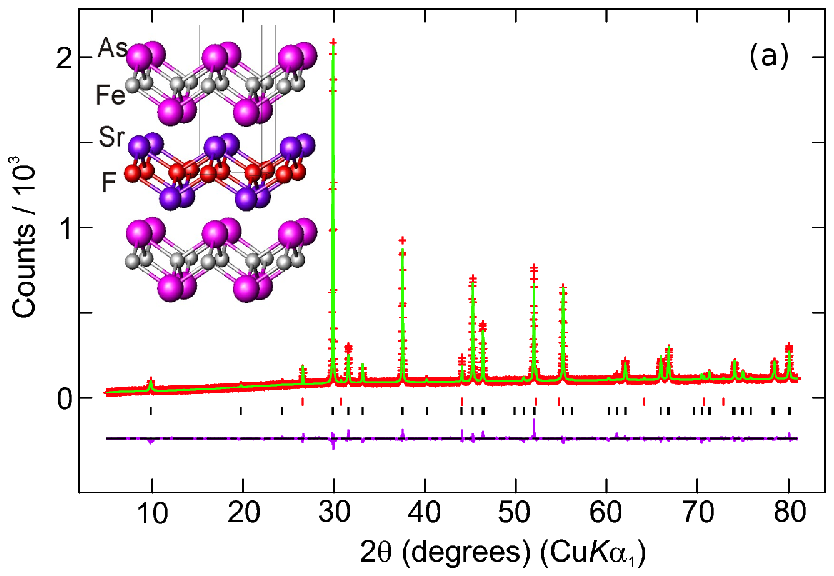}
\includegraphics[width=8.5cm]{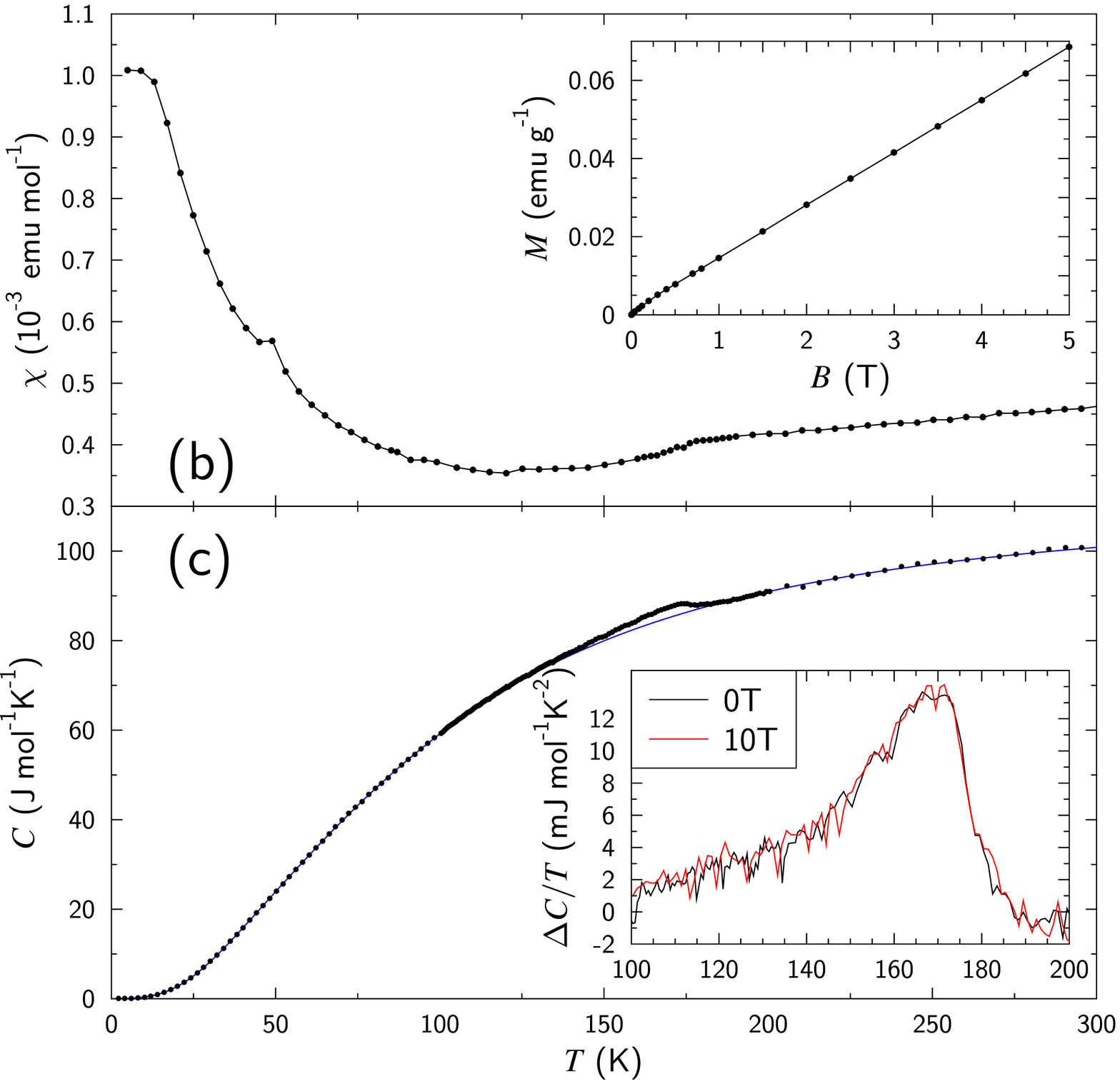}
\caption{ 
(Color online.) 
(a) Rietveld refinement against powder X-ray diffraction data, 
$\chi^2 = 1.98$, wRp $= 0.044$. The inset shows the structure.
(b) DC magnetization measurements vs. temperature in a field of 0.1~T. 
The small feature at about 50K is probably due to a small amount of 
adsorbed O$_2$ apparent because of the small sample moment.
(Inset) Magnetization vs. field at $300$~K.
(c) Heat capacity $C(T)$ showing the peak at $T_s$ which is highlighted 
in the inset. The line shows the lattice heat capacity fit discussed 
in the text. 
\label{bulk}}
\end{figure}

The SrFeAsF sample was synthesized in a two step process similar to that 
described in Ref.~\onlinecite{tegel08arxiv}. Stoichiometric quantities of sublimed 
strontium metal (Alfa 99.9~\%), strontium fluoride powder (Alfa 99.9~\%), iron powder 
(Alfa, 99.998~\%), and arsenic pieces (Alfa, 99.9999~\%; ground into powder) were 
ground together and sealed in a 9 mm diameter niobium tube. This was heated at 
1$^{\circ}/$min to 500$^{\circ}$C and this temperature was maintained for 
12 hours to ensure complete reaction of the volatile components before heating 
at 1$^{\circ}/$min to 900$^{\circ}$C. After 40 hours at 900$^{\circ}$C 
the product was removed from the Nb tube, ground to a fine powder, pressed into 
a pellet, and placed into an alumina crucible which was then sealed in a 
pre-dried evacuated silica tube. This was heated at 1$^{\circ}$C/min to 
1000$^{\circ}$C for 48 hours and then cooled at the natural rate of the 
furnace to room temperature. All manipulation was carried out in an argon-filled 
glove box. Analysis of the product by laboratory X-ray powder diffraction 
(PANAlytical X-pert PRO) [Fig~\ref{bulk}(a)] revealed that the sample consisted of 
about 97~\% by mass SrFeAsF; SrF$_2$ was identified as a crystalline impurity phase, 
but no other crystalline binary or ternary impurity phases were identified. The 
refined room temperature lattice parameters of SrFeAsF were $a = 4.00059(3)$~\AA, 
$c = 8.9647(1)$~\AA, $V = 143.478(4)$~\AA$^3$ consistent with other 
reports.~\cite{tegel08arxiv} Measurement of the dc susceptibility was carried 
out in a Quantum Design MPMS5 instrument [Fig.~\ref{bulk}(b)]. The magnetization of
the sample as a function of field at 300 K showed no significant level of 
ferromagnetic impurity. Measurements as a function of temperature in an applied 
field of 1000 Oe revealed very similar behaviour to that reported 
previously.~\cite{tegel08arxiv} A broad feature at around 175 K is consistent 
with the closely associated antiferromagnetic ordering and structural phase 
transitions which occur in related 
compounds.~\cite{delacruz08nature,rotter08prl,sasmal08prl}
Heat capacity measurements were carried out using a Quantum Design Physical 
Properties Measurement System (PPMS) using a standard relaxation time 
approach. A small part of the sample used for $\mu$SR measurements was 
attached to the sample platform using Apiezon N-grease. Measurements were 
corrected for the heat capacity of the sample platform and grease.
Muon-spin rotation ($\mu$SR) experiments~\cite{blundell99} were performed using 
the General Purpose Surface-Muon Instrument (GPS) at the Swiss Muon Source (Paul
Scherrer Institute, Switzerland).  
The measured parameter is the time-dependent muon decay asymmetry, $A(t)$, 
recorded in positron detectors on opposite sides of the sample.
Our sample was a pressed powder pellet of 1~cm diameter mounted inside a 
silver packet on a silver backing plate. This arrangement gives a time and 
temperature independent background to the signal which is straightforward 
to subtract. 

The heat capacity measurements shown in Figure~\ref{bulk}(c) show a clear feature 
at the structural transition and no anomalies or effects due to latent heat were 
evident at any other temperatures. Our data are in good agreement with those reported 
on this compound by Tegel~{\em et al.}~\cite{tegel08arxiv}. To separate the lattice 
and magnetic contributions to the heat capacity, we estimated the lattice 
background using the function:
\begin{equation}
C (T) = \gamma T + A_{\rm D}C_{\rm D}(T,\theta_{\rm D}) + A_{\rm E}C_{\rm E}(T,\theta_{\rm E}),
\label{hcfit}
\end{equation}
where $\gamma$ is the Sommerfeld coefficient, and $C_{\rm D}$ and $C_{\rm E}$ 
are Debye and Einstein terms respectively. This was found to be an effective 
model for oxypnictides in Ref.~\onlinecite{baker08njp}. The parameters extracted 
from this fit (excluding data between $100$ and $185$~K) were 
$\gamma = 3.44(7)$~mJmol$^{-1}$K$^{-2}$, 
$A_{\rm D} = 56.6(5)$~Jmol$^{-1}$K$^{-1}$, 
$\theta_{\rm D} = 237(1)$~K,
$A_{\rm E} = 52.2(4)$~Jmol$^{-1}$K$^{-1}$, and 
$\theta_{\rm E} = 407(3)$~K. 
These are comparable with the values determined for oxypnictide materials without 
rare-earth magnetic moments.~\cite{mcguire08prb,baker08njp} 
The magnetic contribution is plotted in the inset to Figure~\ref{bulk}(c)
showing that zero-field and 10~T measurements were effectively identical, 
and the integrated magnetic entropy is $0.5$~Jmol$^{-1}$K$^{-1}$. While this is 
a small entropy change, it is twice the value observed in LaFeAsO, where features 
at both the structural and magnetic transitions are evident.~\cite{mcguire08prb} 
SrFe$_2$As$_2$ has a far larger entropy change at the combined first-order 
structural and magnetic transition, $\sim 1$~Jmol$^{-1}$K$^{-1}$. 
The majority of the entropy change in SrFeAsF occurs close to $T_s = 175$~K, but 
it appears that another much broader feature at lower temperature also contributes. 
Since we find long-range magnetic ordering at $T_{\rm N} = 120$~K using $\mu$SR 
(described below), it seems that the broad feature is likely to have a magnetic 
origin. The lack of a distinct anomaly in the specific heat (or in magnetization 
or resistivity data),~\cite{tegel08arxiv} suggests that the residue below the 
structural transition comes from the build up of 2D correlations within the FeAs planes. 
Gaining a rough estimate of the in-plane exchange constant $J \sim 250-300$~K from the 
position of the hump, and knowing $T_{\rm N} = 120$~K, we can estimate the out-of-plane 
exchange constant $J_{\perp} \sim 0.05 J$, consistent with the lack of any observed 
anomaly at $T_{\rm N}$.~\cite{sengupta03,yasuda05} This is a similar situation to 
that in La$_2$CuO$_4$,~\cite{sun91} though with a lower anisotropy in the exchange 
constants and a smaller separation between the structural and magnetic transitions. 

\begin{figure}[t]
\includegraphics[width=8.5cm]{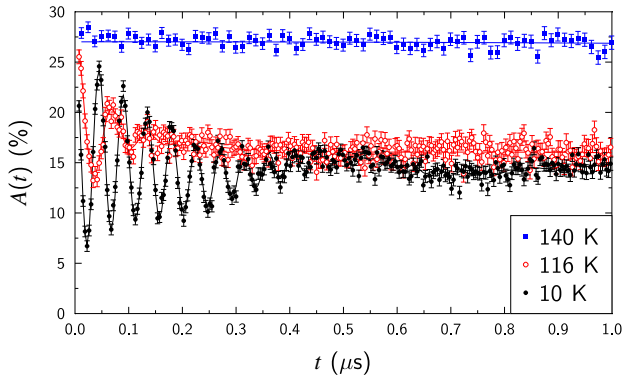}
\caption{
(Color online.) 
Muon asymmetry data for \chem{SrFeAsF} showing the spin precession signal
evident at low-temperature, the greater damping of the oscillations close to 
$T_{\rm N}$, and the paramagnetic signal at $140$~K. The data are fitted to 
Eq.~\ref{fitfunc} with the parameters shown in Figure~\ref{parameters}.
\label{rawdata}}
\end{figure}
In Figure~\ref{rawdata} we present muon decay asymmetry data at temperatures 
of 10, 116, and 140~K. At low temperatures, up to around 75~K, two oscillations 
are clearly resolved but as we approach $T_{\rm N}$ the broadening of each 
of the oscillations grows until they are both overdamped. This overdamped 
behavior is seen in the 116~K data set. Immediately above the magnetic 
ordering transition the muon decay asymmetry takes the exponential form 
expected for a paramagnet with electronic fluctuations faster than the 
characteristic time of the measurement. The data set at 140~K shown in 
Figure~\ref{rawdata} is very similar to all those taken above the magnetic 
ordering temperature, and we saw no change in the relaxation signal when 
passing through $T_s = 175$~K.

\begin{figure}[t]
\includegraphics[width=8.5cm]{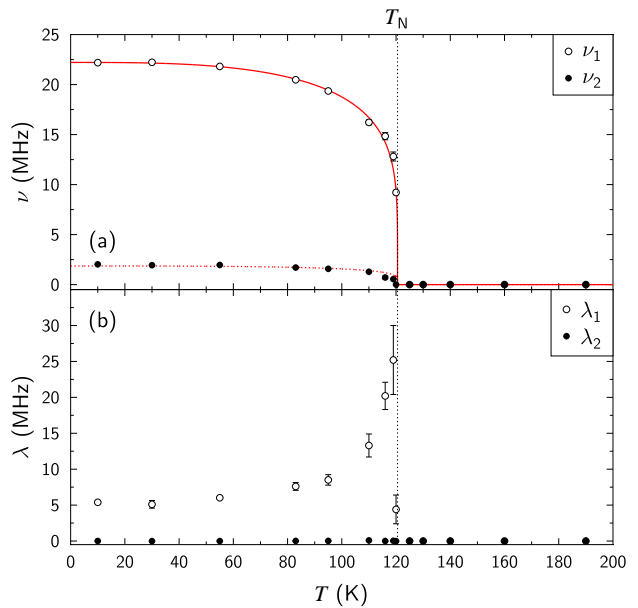}
\caption{
(Color online.) 
Parameters extracted from fitting raw asymmetry data using Eq.~\ref{fitfunc} 
described in the text. 
(a) Oscillation frequencies $\nu_1$ and $\nu_2$ with lines drawn showing the 
power law function described in the text. It is noticeable that the sharp drop-off 
in the frequencies near to the transition is poorly described by this function.
(b) Linewidths of the two oscillating components, $\lambda_1$ and $\lambda_2$.
\label{parameters}}
\end{figure}
Observing two precession frequencies in the magnetically ordered phase 
and finding that the temperature dependent relaxation is Gaussian 
(suggesting that the fluctuations of electronic moments are motionally 
narrowed), we were able to describe the raw asymmetry data using the 
fitting function:
\begin{equation}
A(t) =
\sum^{1,2}_{i} A_{i} e^{-\lambda_{i}t}\cos(2\pi\nu_{i}t)
+ A_3 e^{-\sigma^{2}t^{2}} + A_{bg} e^{-\Lambda t}.
\label{fitfunc}
\end{equation}
The first two terms describe two damped oscillations, the third term 
describes the Gaussian relaxation for muon spins with their direction 
along that of the local field at their stopping site, which are 
depolarized by a random distribution of nuclear moments, and the final 
term describes the weak temperature-independent depolarization observed 
for muons stopping outside the sample. 
Above the magnetic ordering transition there is no oscillatory signal and 
we set $A_1 = A_2 = 0$. 
In many fluorine containing magnets a characteristic signal due to the formation 
of a bound state between a positive muon and one or more fluoride ions is 
observed above the magnetic ordering transition.~\cite{brewer86prb,lancaster07prl}
No such signal is observed in SrFeAsF, probably because the magnetic 
ordering transition is at too high a temperature for the muons to be sufficiently 
well bound.

The parameters derived from fitting Eq.~\ref{fitfunc} to the raw data are shown 
in Figure~\ref{parameters}. The two precession frequencies plotted in 
Figure~\ref{parameters}(a) are well defined and at low-temperature appear to follow 
a conventional power law. Fitting the upper precession frequency to the function 
$\nu(T)~=~\nu(0)(1-(T/T_{\rm N})^{\alpha})^{\beta}$ leads to
$T_{\rm N}~=~120.6(3)$~K, $\alpha = 3.1(3)$, and $\beta = 0.20(2)$. This is a 
much sharper magnetic transition than in LaFeAsO~\cite{klauss08prl,carlo08arxiv} 
and this would suggest that the magnetism is more two-dimensional in this 
fluoropnictide. Also, $\beta$ is between the values expected for 2D Ising and 
2D XY order parameters, though the sharp drop in the frequencies near to $T_{\rm N}$ 
may mean that this fitting function is less effective in estimating the true critical 
parameters. The higher precession frequency tends to $\nu_{1}(0) = 22.22(5)$~MHz 
and, assuming the same power law, the lower precession frequency tends to 
$\nu_{2}(0) = 1.9(1)$~MHz. These frequencies are a little lower than in 
LaFeAsO~\cite{klauss08prl,carlo08arxiv} but in a similar proportion. 
This suggests the magnetic structure is very similar to LaFeAsO and the ordered Fe 
moments $\mu_{\rm Fe} \sim 0.3~\mu_{\rm B}$.~\cite{delacruz08nature,klauss08prl}
Seeing the lower frequency signal persisting all the way to the magnetic ordering 
transition as Carlo {\em et al.}~\cite{carlo08arxiv} did in LaFeAsO suggests 
that this minority oscillation signal is intrinsic to the sample, and reflects the 
antiferromagnetic structure being sampled at a different site within the structure.
It had previously been suggested that the some magnetic signals in these pnictide 
materials originated in FeAs impurities (e.g.~Ref.~\onlinecite{mcguire08prb}) 
but we can discount this possibility for our SrFeAsF sample on the basis of 
$\mu$SR measurements on FeAs and FeAs$_2$, both of which give significantly 
different signals.~\cite{baker08prb} 
In the ordered phase the higher frequency oscillation 
accounts for about $85$~\% of the oscillating amplitude. This amplitude 
ratio for the two oscillating components is similar to the situation in 
\chem{LaFeAsO}, as is the lower frequency signal becoming overdamped 
close to the magnetic ordering transition.~\cite{klauss08prl,carlo08arxiv} 
The linewidths $\lambda_1$ and $\lambda_2$ [shown in Figure~\ref{parameters}(b)] 
are both much smaller than the respective precession frequencies at low temperatures, 
giving rise to the clear oscillations seen in the 10~K data in Figure~\ref{rawdata}, 
and then grow towards the ordering transition giving the overdamped oscillations 
seen in the $116$~K data. 

Our results have shown that long ranged, three-dimensional antiferromagnetic 
ordering in SrFeAsF occurs, but with a greater separation between the structural 
and magnetic ordering transitions ($T_s - T_{\rm N} \sim 50$~K) than in 
comparable oxypnictide compounds (e.g. LaFeAsO). While the $\mu$SR measurements 
show that the magnetic environment within the FeAs planes is very similar to 
that in oxypnictide compounds, we note that the magnetic ordering transition is 
not as clear in the magnetization and heat capacity measurements. The heat capacity 
and $\mu$SR measurements, in particular the lack of a heat capacity anomaly at 
$T_{\rm N}$ and the low value of $\beta = 0.2$, both suggest far more two-dimensional 
magnetic interactions than in oxypnictide compounds, consistent with the increased 
separation $T_s - T_{\rm N}$. This is also consistent with the expectation that the 
interplanar exchange mediated by a fluoride layer will be weaker than that 
mediated by an oxide layer.

Part of this work was performed at the Swiss Muon Source, Paul
Scherrer Institute, Villigen, CH. We are grateful to Alex Amato
for experimental assistance and to the EPSRC (UK) for financial
support.



\begin{thebibliography}{xx}
\bibitem{manousakis91}
E.~Manousakis, Rev.\ Mod.\ Phys. {\bf 73}, 1 (1991).
\bibitem{kastner98}
M.~A.~Kastner {\em et al.},
 {Rev.\ Mod.\ Phys.} {\bf 70}, 897 (1998).
\bibitem{sengupta03}
P.~Sengupta {\em et al.}, 
Phys.\ Rev.\ B {\bf 68}, 094423 (2003).
\bibitem{yasuda05}
C.~Yasuda {\em et al.}, 
Phys.\ Rev.\ Lett. {\bf 94}, 217201 (2005).
\bibitem{kamihara08}
Y. Kamihara {\it et al.},
J. Am. Chem. Soc. {\bf 130}, 3296 (2008).
\bibitem{zhao08nmat}
J. Zhao {\em et al.}, 
Nature Materials {\bf 7}, 953 (2008).
\bibitem{tegel08arxiv}
M.~Tegel {\em et al.}, 
\eprint{arXiv:0810.2120} (unpublished).
\bibitem{han08prb}
F.~Han {\em et al.}, 
Phys.\ Rev.\ B {\bf 78}, 180503(R) (2008).
\bibitem{matsuishi08jacs}
S.~Matsuishi {\em et al.}, 
{J.\ Am.\ Chem.\ Soc.} {\bf 130}, 14428 (2008).
\bibitem{wu08arxiv}
G. Wu {\em et al.}, 
\eprint{arXiv:0811.0761} (unpublished).
\bibitem{matsuishi08arxiv}
S.~Matsuishi {\em et al.}, 
\eprint{arXiv:0811.1147} (2008).
\bibitem{matsuishi08jpsj}
S.~Matsuishi {\em et al.}, 
{J.\ Phys.\ Soc.\ Jpn.} {\bf 77}, 113709 (2008).
\bibitem{zhu08arxiv}
X. Zhu {\em et al.}, 
\eprint{arXiv:0810.2531} (unpublished).
\bibitem{delacruz08nature}
C.~{de la Cruz} {\em et al.}, 
Nature {\bf 453}, 899 (2008).
\bibitem{chen08prb}
Y.~Chen {\em et al.},
 {Phys.\ Rev.\ B} {\bf 78}, 064515 (2008).
\bibitem{kimber08prb}
S.~A.~J.~Kimber {\em et al.}, 
 {Phys.\ Rev.\ B} {\bf 78}, 140503(R) (2008).
\bibitem{mcguire08prb}
M.~A.~McGuire {\em et al.}, Phys.\ Rev.\ B {\bf 78}, 094517 (2008).
\bibitem{mcguire08njp}
M.~A.~McGuire {\it et al.}, \eprint{arXiv:0811.0589} (unpublished).
\bibitem{luetkens08arxiv}
H. Luetkens {\it et al.},
\eprint{arXiv:0806.3533} (2008).
\bibitem{drew08arxiv}
A. J. Drew {\it et al.},
\eprint{arXiv:0807.4876} (unpublished).
\bibitem{krellner08prb}
C. Krellner {\em et al.},
%
Phys.\ Rev.\ B {\bf 78}, 100504(R) (2008).
\bibitem{rotter08prl}
M. Rotter {\it et al.},
Phys. Rev. Lett. {\bf 101}, 107006 (2008).
\bibitem{sasmal08prl}
K. Sasmal {\it et al.},
Phys. Rev. Lett. {\bf 101}, 107007 (2008).
\bibitem{blundell99}
S. J. Blundell, Contemp. Phys. {\bf 40}, 175 (1999).
\bibitem{baker08njp}
P.~J.~Baker {\em et al.},
\eprint{arXiv:0811.2494} (unpublished).
\bibitem{sun91}
K.~Sun {\em et al.},
%
Phys.\ Rev.\ B {\bf 43}, 239 (1991).
\bibitem{brewer86prb}
J. H. Brewer {\em et al.}, Phys. Rev. B {\bf 33}, 7813 (1986).
\bibitem{lancaster07prl}
T. Lancaster {\em et al.},
Phys.\ Rev.\ Lett. {\bf 99}, 267601 (2007).
\bibitem{klauss08prl}
H.-H. Klauss {\it et al.},
Phys. Rev. Lett. {\bf 101}, 077005 (2008).
\bibitem{carlo08arxiv}
J. P. Carlo {\it et al.},
\eprint{arXiv:0805.2186} (unpublished).
\bibitem{baker08prb}
P.~J.~Baker {\em et al.}, 
 {Phys.\ Rev.\ B} (accepted) \eprint{arXiv:0809.2522} (2008).
\end{thebibliography}
\end{document}